\documentclass[pra, 12 pt]{revtex4}
\usepackage[english]{babel}
\usepackage{amsmath}
\usepackage{epsfig}

\begin{document}

%\draft

\title{\bf JEANS INSTABILITY AND ANTI-SCREENING IN GRAVITATIONAL-ANTIGRAVITATIONAL MODEL OF UNIVERSE}

\author{I. Gribov $^1$, S.A. Trigger$^{2,3}$}
\address{$^1$ Institute for the Time Nature Exploration,
MSU - 119991, Moscow, Russia\\
e-mail: igribov@aol.com;\\
$^2$ Joint\, Institute\, for\, High\, Temperatures, Moscow\, 125412, Russia;\\
e-mail: satron@mail.ru\\
$^3$ Eindhoven  University of Technology, P.O. Box 513, MB 5600
Eindhoven, The Netherlands}

\begin{abstract}
The hypothesis of antigravitational interaction of elementary particles and antiparticles
is considered on the basis of the simple two-component hydrodynamic model
with gravitational repulsion and attraction. It is shown increasing of the Jeans instability
rate, the presence of antiscreening and the dominative role of the gravitational
repulsion as a possible mechanism for spatial separation of matter and antimatter in
Universe, as well as the observable acceleration of the far galaxies. The sound wave
is found for the two-component gravitational-antigravitational system, which starts
for k = 0 in the case of annihilation neglecting. The suggested approach permits to
reestablish the idea about baryon symmetry of Universe, causing its steady flatness
of the large scale and accelerated Universe expansion.

\end{abstract}

\maketitle

\section{Introduction}

Isaac Newton (in his letter to Richard Bentley, 1692) first
suggested that self-gravity in infinite Universe would lead to the
observed mass distribution [1]. First
quantitative description of fragmentation of matter due to
self-gravity has been done by Jeans [2] in 1902. According [2] self-gravitating infinite
uniform gas at rest should be unstable against small perturbations
proportional to $\exp [i ({\bf kr} - \omega t)]$. Linearization of
equations of ideal hydrodynamics and Poisson equation for the
gravitational potential results in the dispersion equation
\begin{equation}
\omega ^2 = c^2 k^2 - \Omega^2, \label{om}
\end{equation}
where $\Omega= (4 \pi G \rho)^{1/2}$\, is the Jeans gravitational
frequency, $\rho$ is the density, $c=(\gamma T/m)^{1/2}$  is the
adiabatic sound velocity, $\gamma = 5/3$ is the ratio of specific
heats, $T$ is the gas temperature in the energy units, $m$ is the
particle mass, and $G$ is the gravitational constant.

As seen from Eq. (1), $\omega ^2$ becomes negative, and the instability
arises when the perturbation wavelength, $\lambda = 2\pi /k$
exceeds the critical value:
\begin{eqnarray}
\lambda > \lambda _c = c \sqrt {\frac{\pi}{G\rho}}. \label{lam}
\end{eqnarray}

The pressure gradient tends to quench the instability. This force
dominates over the gravity force resulting in stabilization with
$\lambda \leq \lambda _c$. Thus, originally uniform gas, due to
the instability, should break into clots with characteristic size
of the order of $\lambda _c$. It is worthy of mentioning that
$k^{-1}_{c} = \lambda _c /2\pi = c/(4\pi G\rho)^{1/2}$ is the only
characteristic linear scale, which may be constructed of the
parameters inherent for the problem under consideration.

Kinetic theory of the Jeans instability was given in [3-6] using
methods of plasma physics. It is worthy of mentioning that more realistic models, like
expanding Newtonian world-model [7], lead to the same Jeans
instability criterion (2). In fact, very similar results have been
obtained in [8] by using Friedmann solution for expanding
Universe.

The $\Lambda$-CDM standard model of Big Bang cosmology includes Einsteinian cosmological constant $\Lambda$, hypothetically associated with repulsive dark energy and Cold Dark Matter (CDM). As is known the $\Lambda$-CDM provides an excellent parametrization of the present cosmological data, including the observed perfect flatness and recently discovered accelerating expansion of our Universe on its large scale. However, this model includes two vastly dominating, but unknown and  unobserved components, dark matter and dark energy, that fill more than 95 percents of Universe and the hypothesis of initial hyperinflation, providing surprising flatness of our Universe, including its earlier stages directly after Big Bang. Taking into account this unsatisfactory situation, some authors arrive at an unconventional matter-antimatter symmetric cosmology, in which matter and antimatter gravitationally repulse each other, mutually survived and are always equally presented in Universe, providing zero gravity mass density on the large scale. So, antimatter (with positive inertial mass) is supposed to present a negative active gravitational mass (like the opposite electrostatic charges in plasmas) for all particles-antiparticles, e.g., neutron-antineutron, electron-positron, proton-antiproton etc. The first discussion, related with such hypothesis has been done by L. Schiff [9] (see also, references therein). Later on the different aspects of this idea
have been discussed (see,e.g., [10,11]). In particular, the criticism, concerning the idea of anti-gravitational interaction of matter and antimatter is considered in detail and in many points rejected in [12,13]. Some essential features of the respective cosmology
(the Dirac-Milne one) have been considered recently ([14,15] and references therein).

Accepting the idea of particle-antiparticle gravitational repulsion one can explain the accelerating expansion of the Universe observed on the distances 5-8 billions of the lightyears
and provide the physically clear basis for the concept of dark matter and dark energy
[16,17].

In this paper on the basis of a simple model of two-component gravitational-antigravitational
electrically neutral particles - $"2.G"$ system (e.g., gravitationally neutral neutron-antineutron system) we consider the generalization of the Jeans instability. We demonstrate also that in
contrast with the neutral system of electrical charges (plasma like systems) the neutral system of gravitational charges possesses "anti-screening" property. The remarkable features of gravitational-antigravitational system lead to new opportunity to solve the problem of baryon asymmetry of Universe by recovering the requirement of baryon symmetry. The background of this
symmetry is the spatial separation of the mass of the opposite signs during the early stages
of the Universe evolution. We shortly discuss this effect in the framework of the considering
simple model.

\section{Modes and instability for two component gravitational hydrodynamics with annihilation}

Let us start from hydrodynamic equations for two species of electrically neutral particles (e.g., neutrons and antineutrons), assuming that their "gravitational charges"\, are opposite and the gravitational interaction of the particles and antiparticles is repulsive
\begin{eqnarray}
 \frac{\partial n_a}{\partial t} + div (n_a {\bf V}_a)=- \nu_{a,\bar{a}} n_a n_{\bar{a}}, \label {A1}
\end{eqnarray}
\begin{eqnarray}
m n_a [\frac{\partial {\bf V}_a}{\partial t} + ({\bf V}_a\cdot\nabla) {\bf V}_a] = -\frac{\partial p_a}{\partial {\bf r}}+n_a \textbf{F}_a\label{lam 2},\, \textbf{F}_a=-m_a \nabla \Phi. \label {A2}
\end{eqnarray}
Here $n_a$ is the density of particles of specie $a$, $p_a=n_a({\bf r},t)T$ is the respective pressure, index $a=\{m, -m \}$ and $\bar{a}=-a$. The gravitational masses have the opposite signs (for certainty $m>0$), however, the inertial mass is always positive and equals $\mid m_a \mid=m$. We suppose that temperature $T$ is a constant and omit the energy equation, neglecting of photons creation in the annihilation process. The value $\nu_{a,\bar{a}}=\nu_{\bar{a},a}\equiv \nu$ is the characteristic rate for the particle-antiparticle annihilation process.

The Poisson equation reads
\begin{eqnarray}
\Delta \Phi = 4\pi G m (n_{m}-n_{-m}).\label {A3}
\end{eqnarray}
After linearization of Eqs. (3)-(5) on small deviations from the homogeneous state $n_m^0=n_{-m}^0$, $V_{m}^0=V_{-m}^0=0$ and $\Phi=0$ we arrive to the dispersion relation
\begin{eqnarray}
\omega^4+i 2\omega^3 n^0\nu + 2 \omega^2  (\Omega^2-c^2 k^2)\nonumber\\+
i\omega 2 n^0\nu(2 \Omega^2- c^2 k^2)+c^4 k^4-2 c^2\Omega^2 k^2=0, \label {A4}
\end{eqnarray}
where the following notations are used $c^2=T/m$, $\Omega^2=4\pi G m n^0$.
For $k=0$ Eq. (6) reads
\begin{eqnarray}
\omega^4+ 2 i  \omega^3 n^2\nu+ 2 \omega^2 \Omega^2+4 i\omega  n \nu \Omega^2=0.  \label {A5}
\end{eqnarray}

The explicit nontrivial solutions of Eq. (7) are
\begin{eqnarray}
\omega_{1,2}=\pm i \sqrt 2 \,\Omega,\;  \omega_3=-2i\nu n. \label {A6}
\end{eqnarray}
The unstable solution $\omega_{2,3}=i \sqrt 2 \,\Omega$ is similar to the Jeans instability in the system of gravitational masses, however, the rate of this "balling"\, instability is higher in $\sqrt 2$ times than the Jeans one due to the presence of the subsystem of masses with the opposite gravitational charge.

In the recent work Chris Collins [18] considered the so-called "Planck-cluster problem" and estimated the number of massive clusters in the Universe. According to the estimation this number is about a factor of two below the prediction based on the Planck CBR analysis. From our point of view this estimation can be connected with the factor $\sqrt 2$ in the above instability. The obtained instability rate confirms the assumption [12] that gravitational repulsion of a matter - antimatter system accelerates consolidation of matter and antimatter clusters.

These "balls"\, of the opposite gravitational charge continue to repulse one from another providing the accelerated expansion of the whole system. The third solution describes the purely damping annihilation mode.

For the case of the model with $\nu=0$ the solutions Eq. (6) for finite $k$ reads
\begin{eqnarray}
\varpi_1^2=-2\Omega^2+c^2 k^2, \qquad \varpi_2^2=c^2k^2.  \label {A7}
\end{eqnarray}
The peculiarity of the "balling"\, instability is stabilization on the smaller characteristic size than the Jeans instability. The second essential peculiarity of the system with antigravitation is the appearance of the sound (linear on $k$) mode. It is easy to see that the sound mode for a small, but finite rate of annihilation has the damping spectrum
\begin{eqnarray}
\varpi_3=c k - i n^0 \nu. \label {A8}
\end{eqnarray}
Therefore, the "sound wave" exists only for $k \gg k_0=n^0 \nu/c$. For the opposite inequality $k \ll k_0=n^0 \nu/c$ this mode is absent, according to (8).

\section{Two-component gravitational system: anti-screening of an immobile gravitational charge}

For the case of the point gravitational mass $\mu_0$ (providing the spherical symmetry) in a two-component infinite gravitational system with immobile (${\bf V}_e={\bf V}_i=0$) masses $m_{+}=-m_{-}=m$ the stationary distribution at $T=const$ describes by equations
\begin{eqnarray}
T \frac{\partial n_m}{\partial r}=-m n_m \frac{\partial \Phi(r,t)}{\partial r} \qquad \rightarrow  \qquad n_m=n_{m,0} \exp (-m \Phi /T)\nonumber\\
T \frac{\partial n_{-m}}{\partial r}=m n_{-m} \frac{\partial \Phi(r,t)}{\partial r} \qquad \rightarrow  \qquad n_{-m}=n_{-m,0} \exp (m \Phi /T);\label {B1}
\end{eqnarray}
\begin{eqnarray}
\triangle \Phi(r)=4 \pi G [m n_m-m n_{-m})+4 \pi G \mu_0 \delta (\textbf{r})\nonumber\\=4 \pi G [m n_{m,0}\exp (- m\Phi /T) -m n_{-m, 0} \exp (m\Phi /T)]+4 \pi G \mu_0 \delta (\textbf{r}).\label {B2}
\end{eqnarray}
Let us suppose (as in the case of an electrically charged particles in quasineutral weakly non-ideal plasma) that linearization is possible. For the case $n_{m,0}=n_{-m,0}=n_0$, introducing the value of gravitational radius $R_G^2=T/4 \pi G m^2 n_0$ we arrive at equation
\begin{eqnarray}
\triangle \Phi(r)+ 2 \Phi /R_G^2=4 \pi G \mu_0 \delta (\textbf{r}).\label {B4}
\end{eqnarray}
For the Fourier-component $\Phi(\textbf{k})=\int d \textbf{r} \exp (-i \textbf{k  }\cdot\textbf{ r}) \Phi(\textbf{r})$
\begin{eqnarray}
-k^2\Phi(k) +\frac{2}{R_G^2}\Phi(k)=4\pi G \mu_0 \delta (\textbf{k}),\label {B5}
\end{eqnarray}
\begin{eqnarray}
\left[-k^2+\frac{1}{R'^2_G}\right]\Phi(k) =4\pi G \mu_0, \label {B6}
\end{eqnarray}
where $2/R_G^2 \equiv 1/R'^2_G$.\label {B7}
\begin{eqnarray}
 \Phi (\textbf{r})=
 -\frac{\mu_0 G }{ r} cos (\frac{r}{R'_G}), \label {B7a}
\end{eqnarray}
\begin{eqnarray}
n_m=n_0 \exp (-m \Phi /T)\rightarrow n_m=n_0 \exp [\frac{m \mu_0 G }{T r} cos (\frac{r}{R'_G})]\simeq n_0  [1+\frac{m \mu_0 G }{T r} cos (\frac{r}{R'_G})],\label {B8}
\end{eqnarray}
\begin{eqnarray}
n_{-m}=n_0 \exp (m \Phi /T)\rightarrow n_{-m}=n_0 \exp [-\frac{m \mu_0 G }{T r} cos (\frac{r}{R'_G})]\simeq n_0  [1-\frac{m \mu_0 G }{T r} cos (\frac{r}{R'_G})]. \label {B9}
\end{eqnarray}
The distributions are crucially different from the two-component plasma. If the signs of the gravitational masses $m$ and $\mu_0$ are the same
the density of masses $m$ increases around $\mu_0$, whereas the density of masses $m_{-}$ decreases around $\mu_0$. Both densities tends to $n_0$ where $r\rightarrow \infty$. Since the density $n_m (r\rightarrow 0)\rightarrow \infty$ linearization is not valid for the problem in general. However, due to linearization it is easy to understand that physical picture in gravitational-antigravitational two-component system (which can be determined as anti-screening) is crucially different from the plasma case. The potential $\Phi$ has the oscillatory behavior.

The attraction of the masses of the same gravitational charge leads to the creation of
the massive clouds of the same gravitational mass (gravitational clusters of particles and
antiparticles), whereas the clouds of the opposite gravitational mass tend to break up.
This trend can play a fundamental role for the global astrophysical processes.

\section{Two-component gravitational system: force}

In the previous sections we considered the processes in initially homogeneous infinite $2.G$ system. Let us consider now forces in the finite initially inhomogeneous $2.G$ system.

After integration of the Poisson equation for the case of spherical symmetry $\triangle \Phi=4 \pi G m ( n_m(\textbf{r})-n_{-m}(\textbf{r}))$, where $\triangle =\frac{1}{r^2}\frac{\partial }{\partial r} (r^2 \frac{\partial\Phi(r)}{\partial r})$. The force on one particle equals $\textbf{F}_m=-m \nabla \Phi$ and $\textbf{F}_{-m}=m \nabla \Phi$
\begin{eqnarray}
\frac{\partial \Phi}{\partial r}-\frac{4\pi G m }{r^2}\int_0^r d r'  r'^2 (n_m(r')- n_{-m}(r')) =0\nonumber\\
\textbf{F}_{m,r}(r)=-m  \frac{\partial \Phi(r)}{\partial r}=-\frac{4\pi G m^2}{r^2}\int_0^r d r'  r'^2 (n_m(r')-n_{-m}(r')).\label {C1}
\end{eqnarray}
For a homogeneous distributions both components inside a sphere radius $R_0$ the force acting on one "$e$" particle ($r$-component) placed on the surface of the sphere
\begin{eqnarray}
\textbf{F}_{m,r}(R_0)==-\frac{4\pi G m^2}{R_0^2}\int_0^{R_0} d r'  r'^2 (n_m(r')- n_{-m}(r'))=-\frac{4\pi G m^2 ((N_m-1)- N_{-m})}{R_0^2}.
\label{C2}
\end{eqnarray}

For $N_m=N_{-m}$ the force is positive $\textbf{F}_{m,r}(R_0)=\frac{ G m^2}{R_0^2}>0$, and therefore always directed from the center.

For the particles of the type $m_{-}$
\begin{eqnarray}
\textbf{F}_{-m,r}(r)= m  \frac{\partial \Phi(r)}{\partial r}=\frac{4\pi G m^2}{r^2}\int_0^r d r'  r'^2 (n_m(r')-n_{-m}(r'))  \nonumber\\ \textbf{F}_{-m,r}(R_0)==\frac{4\pi G m^2}{R_0^2}\int_0^{R_0} d r'  r'^2 (n_m(r')- n_{-m}(r'))=\frac{4\pi G m^2(m (N_e- (N_i-1))}{R_0^2}.
\label{C3}
\end{eqnarray}
 Since $N_m=N_{-m}$ the force is also the same and positive $\textbf{F}_{-m,r}(R_0)=\frac{ G m^2}{R_0^2}>0$. Therefore, for homogeneous spherically symmetric distribution the force on the surface particle of arbitrary sign of mass is always directed from the center.

However, this calculation seems to primitive since the distortion of density around the probe gravitational charge placed on the sphere $R_0$ does not taken into account. Let us estimate the force more accurately assuming that the potential created by this mass $\mu_0$ on the boarder is of order (16). Since the potential of this charge in vacuum is $\Phi_0(r)=-\mu_0 G/r$, the potential created by the surrounding particles equals $\delta \Phi(r)=\Phi(r)-\Phi_0(r)$ (in this approximation we neglect for estimation the spherical asymmetry violation)
\begin{eqnarray}
\delta \Phi (\textbf{r})=
\frac{\mu_0 G }{ r} [1-cos (\frac{r}{R'_G})]. \label{C4}
\end{eqnarray}
Therefore, the force $\textbf{F}_{\mu_0,r}(r\rightarrow 0)$ affected on the probe charge equals
\begin{eqnarray}
\textbf{F}_{\mu_0,r}(r\rightarrow 0)=\lim_{r\rightarrow 0}\frac{\partial \delta \Phi}{\partial r} \simeq \frac{3 \mu_0^2}{4 R'^2_G}.
\label {C5}
\end{eqnarray}
This force is directed outside the spherical volume (along the unit vector $\textbf{r}/r$) independently on the sign of gravitational charge and provides the accelerating expansion of the system under consideration.

Let us now calculate the force for two spherical homogeneous distributions with the different radii $R_m$ and $R_{-m}$. For certainty take $R_{-m}>R_m$. The "global mass neutrality" requires $n_{m,0} R_m^3=n_{-m,0} R_{-m}^3 =N_m=N_{-m}$. Therefore, $n_{m,0}>n_{-m,0}$. The force $\textbf{F}_{m,r}$ inside $R_m$:
\begin{eqnarray}
\textbf{F}_{m,r}(r<R_m)==-\frac{4\pi G m^2}{r^2}\int_0^{r<R_m} d r'  r'^2 (n_{m,0}(r')- n_{-m,0}(r'))\rightarrow \nonumber\\=-\frac{4\pi G m^2 r (n_{m,0}- n_{-m,0})}{3}<0.
\label{C6}
\end{eqnarray}
This force directed to the center. This means the sort of particles $m$ will constrict to the center.

The force $\textbf{F}_{-m,r}$ inside $R_m$
\begin{eqnarray}
\textbf{F}_{-m,r}(r)=\frac{4\pi G m^2 r(n_{m,0}-n_{-m,0})}{3} >0. \label {C7}
\end{eqnarray}
This force directed outside from the center and the sort of particles $m_{-}$ will pushed outside the sphere $R_m$.

The forces $\textbf{F}_{m,r}$ (however, the $m$ particles are absent for $r>R_m$, but we can determine the force acted on an added probe $m$ particle), as well as the force $\textbf{F}_{-m,r}$ outside $R_m$, but inside $R_{-m}>R_m$:
\begin{eqnarray}
\textbf{F}_{m,r}(R_m<r<R_{-m})==-\frac{4\pi G m^2}{r^2}\int_0^{R_m<r<R_{-m}} d r'  r'^2 (n_{m,0}(r')-n_{-m, 0}(r'))\rightarrow \nonumber\\=-\frac{4\pi G m^2 (n_{m, 0}R_m^3- n_{-m,0}r^3)}{3 r^2}<0.
\label{C8}
\end{eqnarray}
This force directed to the center, since $(n_{m, 0}R_m^3- n_{-m,0}r^3)>0$ for $R_m<r<R_{-m}$.

\begin{eqnarray}
\textbf{F}_{-m,r}(R_m<r<R_{-m})==\frac{4\pi G m^2(n_{m,0}R_e^3-n_{-m,0}r^3)}{3 r^2}>0.
\label{C9}
\end{eqnarray}
This force directed outside the center, since $n_{m,0}R_e^3-n_{-m,0}r^3$ since $R_m<r<R_{-m}$. The particles of the sort $m_{-}$ will pushed out the initial sphere.

Finally, for $r>R_i$
\begin{eqnarray}
\textbf{F}_{m,r}(r>R_i)==-\frac{4\pi G m^2 (n_{m,0}R_e^3- n_{-m,0}R_i^3)}{3 r^2}=0,
\label{C10}
\end{eqnarray}
\begin{eqnarray}
\textbf{F}_{-m,r}(r>R_i)==\frac{4\pi G m^2(n_{m,0}R_e^3-n_{-m,0}R_i^3)}{3 r^2}=0.
\label{C11}
\end{eqnarray}
The gravitational forces acting outside the bigger radius are zero due to the "global mass neutrality" requirement $n_{m,0} R_m^3=n_{-m,0} R_{-m}^3 =N_m=N_{-m}$. The force connected with pressure can be easily included. Pressure prevent to the compression by gravitational force directed to the center and contribute to the gravitational force directed outside the sphere.

The same way can be used to consider the different (in particular smooth) spherically-symmetrical profiles and more complicated non-symmetrical profiles. We see the dominative role of gravitational repulsive force between the initially separated "clouds" of particles with the opposite gravitational charges.

\section{Correlation energy calculation}

Although the homogeneous $2.G$ system is unstable it is interesting to calculate energy and to compare the result with the energy of weakly interacted plasma. The correlation energy can be written in the form [16]
\begin{eqnarray}
E_{corr.}=\frac{1}{2V^2}\sum_{a,b} N_a N_b \int\int u_{a,b}w_{a,b}d V_a dV_b,
\label{D1}
\end{eqnarray}
where the potentials equal $u_{a,b}=-m_a m_b G/r$ (repulsion interaction of the particles with opposite gravitational masses) and $w_{a,b}$ is the pair correlation function. Taking into account these notations and following to the standard procedure applied to Coulomb systems we arrive at equation for $w_{a,b}$
\begin{eqnarray}
\Delta w_{a,b}(\textbf{r})=-\frac{4\pi m_a m_b G}{T}\delta(\textbf{r})-\frac{4\pi m_b G}{T V}\sum_c N_c m_c w_{a,c}(\textbf{r}).
\label{D2}
\end{eqnarray}
The solution of this system of equations can be found in the form $w_{a,b}(\textbf{r})=m_a m_b \omega(\textbf{r})$, where the function $\omega(\textbf{r})$ obeys equation
\begin{eqnarray}
\Delta \omega(\textbf{r})+\frac{1}{R'^2_G} \omega(\textbf{r})=-\frac{4\pi G}{T}\delta(\textbf{r}),
\label{D3}
\end{eqnarray}
where $2 R'^2_G=T/4\pi n_0 m^2 G$ and the equalities $m_{+}=-m_{-}=m>0$, $N_{m}=N_{-m}$, $n_{m,0}=n_{-m,0}=n_0=N_m/V$ are taken into account.
The solution for $\omega(\textbf{r})$ reads
\begin{eqnarray}
\omega(\textbf{r})=\frac{G}{r T} cos (r/R').
\label{D4}
\end{eqnarray}
For $E_{corr.}$ we obtain
\begin{eqnarray}
E_{corr.}=-V \frac{T}{16 \pi R'^3_G} \int_0^\infty  cos \zeta  d\zeta=-V T \frac{(8\pi n_0 m^2 G)^{3/2}}{16 \pi T^{3/2}} \int_0^\infty  cos \zeta  d \zeta,
\label{D5}
\end{eqnarray}
or taking into account the finite value of a large volume ($V=4 \pi R^3/3$)
\begin{eqnarray}
E_{corr.}=-\frac{\pi R^3 T}{12 \pi R'^3_G} \int_0^{R/R'_G}  cos \zeta  d\zeta=- T \frac{(8 \pi n_0 m^2 G)^{3/2}R^3}{12 T^{3/2}} \int_0^{R/R'_G}  cos \zeta  d \zeta.
\label{D6}
\end{eqnarray}
The thermodynamic limit is absent.

\section{Conclusions}

The hypothesis about antigravitational interaction between antiparticles is discussed on the basis of the simplest model of electrically
neutral system of particles and antiparticles (e.g., neutrons and antineutrons). Assuming the existence of antigravitation for antiparticles and the condition of global gravitational neutrality we found the change of the Jeans instability rate. We suppose the connection between this change and the estimation of the number of massive clusters in Universe. It is shown also the presence of
"antiscreening" in the model under consideration. Both effects lead to the conclusion that
not space itself is expanded via hyperinflation, but assumedly matter and antimatter have tendency to spatial separation, performing the observed quasi-linear Hubble-like Universe expansion. The force calculation calculation supports the hypothesis that the mechanism of the observable acceleration of the far located galaxies in Universe can be explained on the basis of the considered antigravitational interaction
between matter and antimatter. Therefore, this interaction can play a role of the dark energy and dark matter in the globally gravitationally neutral Universe.
The essential advantage of the hypothesis is the opportunity to reestablish the baryon symmetry of Universe. The quantity of matter and antimatter in Universe is the same, but due to gravitational - antigravitational of the massive matter-clouds and antimatter-clouds repulsion they separated during the gravitational Universe evolution. Creation of these massive clouds is conditioned bythe gravitational attraction of the particles (as well as antiparticles) of the same specie.
The structure of the observable Universe on a huge distances can be similar to a soap
foam or a huge bubbles, which surfaces consist from the gravitational and antigravitational
objects (e.g., galactic and antigalactic clusters), where an average surface gravity mass density on the large scale remains zero.
The problem of the non-linear time evolution for the
models with antigravitational interaction of antiparticles has to be considered analytically
and numerically, taking into account the transformation of elementary particles and
antiparticles, electromagnetic radiation and creation of the elements on the more late study
of Universe evolution.

The gravitational interaction of matter with antimatter has not been conclusively observed by physicists in laboratories on the Earth. The first enough sensitive and decisive experimental tests of the antihydrogen gravitational properties will be conducted to the end of the next year 2015 on ALPHA [20], AEgIS [21] and GBAR [22] experiments at CERN. \\

\end{document}